\documentclass[12pt]{iopart}

\usepackage{iopams}
\usepackage{color}
\usepackage[english]{babel}

\usepackage[pdftex]{graphicx}

\definecolor{mygrey}{gray}{0.35}
\definecolor{myblue}{rgb}{0.2,0.2,0.8}
\definecolor{myzard}{cmyk}{0,0,0.05,0}
\definecolor{mywhite}{rgb}{1,1,1}
\definecolor{mywhite}{rgb}{1,1,1}
\definecolor{myred}{rgb}{1,0.,0.3}

\usepackage[colorlinks=true,citecolor=myblue,linkcolor=myred]{hyperref}

\def\dd{\mathord{\rm d}}
\def\ee{\mathord{\rm e}}
\def\ii{\mathord{\rm i}}

\renewcommand{\aa}{{\rm a}}

 \newcommand{\ket}[1]{|#1\rangle}
 \newcommand{\bra}[1]{\langle #1|}

 \newcommand{\ud}[1]{{#1^{\dagger}}}
 
\def\oned{\mathrm{1d}}

\def\II{{\rm I}}
\def\DD{{\rm D}}
\def\PP{\mathbb{P}}
\def\QQ{\mathbb{Q}}

\newcommand{\binom}[2]{ {#1 \choose #2} }

\begin{document}

\title[Universal Quantum Computation in Waveguide QED using DFS]{Universal Quantum Computation in Waveguide QED using Decoherence Free Subspaces}

\author{V. Paulisch$^1$,  H. J. Kimble$^{2,3}$ and A. Gonz\'alez-Tudela$^1$}

\address{$^1$ Max-Planck-Institute of Quantum Optics, Hans-Kopfermann-Strasse 1, 85748 Garching, Germany}
\address{$^2$ Norman Bridge Laboratory of Physics 12-33, California Institute of Technology, Pasadena, CA 91125, USA}
\address{$^3$ Institute for Quantum Information and Matter, California Institute of Technology, Pasadena, CA 91125, USA}

\ead{alejandro.gonzalez-tudela@mpq.mpg.de}

\begin{abstract}
The interaction of quantum emitters with one-dimensional photon-like reservoirs induces strong and long-range dissipative couplings that give rise to the emergence of so-called Decoherence Free Subspaces (DFS) which are decoupled from dissipation. When introducing weak perturbations on the emitters, e.g., driving, the strong collective dissipation enforces an effective coherent evolution within the DFS. In this work, we show explicitly how by introducing single-site resolved drivings, we can use the effective dynamics within the DFS to design a universal set of one and two-qubit gates within the DFS of two-level atom-like systems.  Using Liouvillian perturbation theory we calculate the scaling with the relevant figures of merit of the systems, such as the Purcell Factor and imperfect control of the drivings. Finally, we compare our results with previous proposals using atomic $\Lambda$ systems in leaky cavities.
\end{abstract}


\vspace{2pc}

 
\maketitle

\section{Introduction}

Recent theoretical and experimental work has shown that an attractive configuration to engineer strong collective dissipation is given by one-dimensional (1d) photonic-like systems such as photonic crystal waveguides \cite{joannopoulos_book95a,laucht12a,goban13a,yu14a,tiecke14a,sollner14a,arcari14a,goban15a,young15a}, optical fibers \cite{vetsch10a,goban12a,mitsch14a,petersen14a,beguin14a}, metal \cite{chang07a,dzsotjan10a,gonzaleztudela11a,bermudez15a} and graphene plasmonic  \cite{koppens11a,huidobro12a,christensen11a,martinmoreno15a} waveguides or superconducting circuits \cite{mlynek14a}. Their interaction with quantum emitters, usually referred to as waveguide QED, offers interesting characteristics:
\textit{i)} the density of modes of the waveguide is inversely proportional to the group velocity $1/v_g(\omega_{\aa})$, and therefore is strongly enhanced when the atomic frequency is in a region of slow light, e.g., in photonic crystal waveguides close to a band edge. This enhancement implies achieving regions of a large decay rate into the waveguide, $\Gamma_\oned$, compared to other decay channels, denoted by $\Gamma^*$, characterized through the  \emph{Purcell Factor}, $P_\oned=\Gamma_\oned/\Gamma^*$;
\textit{ii)} the 1d guided modes retain a small modal area $\lesssim \lambda_\aa^2$, for propagation lengths $L_{\mathrm{prop}} \gg \lambda_\aa$ (the wavelength of the 1d mode associated to the atomic frequency considered); \textit{iii)} the interaction is \emph{strongly} long-range, favoring individual adressing, and it can even be homogeneous if the positions of the atom-like systems are chosen properly \cite{lehmberg70a,lehmberg70b}, in contrast to 2d or 3d system. This collective dissipation leads to the emergence of subradiant states that form the so-called Decoherence-Free Subspace (DFS) \cite{zanardi97a,lidar98a}. 

Previous works have already considered how to use the DFS of two atoms trapped in leaky cavities to design one and two-qubit gates using three-level $\Lambda$-type schemes \cite{beige00a,beige00b,tregenna02a}, where two atomic hyperfine levels are used to encode the qubit. In the light of the variety of systems available nowadays that allows to engineer robust one-dimensional DFS \cite{joannopoulos_book95a,laucht12a,goban13a,yu14a,tiecke14a,sollner14a,arcari14a,goban15a,young15a,vetsch10a,goban12a,mitsch14a,petersen14a,beguin14a,chang07a,dzsotjan10a,gonzaleztudela11a,bermudez15a,koppens11a,huidobro12a,christensen11a,martinmoreno15a,mlynek14a}, which may couple to different types of quantum emitters, e.g., atoms, quantum dots, NV centers or superconducting qubits, 
it is interesting to revisit the problem and fill some of the gaps that have not been considered so far, namely,:
\textit{i)} how to encode decoherence-free qubits using only two-level systems (TLS) (as $\Lambda$-schemes might not be available for all platforms); 
\textit{ii)} extend the proposal to systems with more than two atoms;
\textit{iii)} analyze the scaling of the fidelities with the relevant figures of merit of the system, e.g., $P_{\oned}$, that in previous works \cite{beige00a,beige00b,tregenna02a} was mainly done through numerical analysis; 
\textit{iv)} discuss the consequences of imperfect addressing on the fidelity of the gates.

In this work, we show an implementation of universal quantum gates by using $N$ TLS strongly coupled to 1d photon-like reservoirs. First, we show that by \emph{pairing} the TLS, we can define decoherence free-qubits in the singlet (i.e., antisymmetric) states of each pair. The combination of these singlets form the so-called \textit{computational subspace} where we define our operations. Then, we explicitly show how to build single qubit (e.g., phase gates and Pauli-$X$ gates) and two-qubit (e.g., controlled $Z$-gates) operations within the computational subspace without coupling to the other states in the DFS. By using Liouvillian perturbation theory, we obtain analytical expressions for the scaling of the fidelities of the operation ($1-F\propto 1/\sqrt{P_\oned}$) and estimate the error when increasing the number of atoms. Finally, we revisit the problem of the implementation with $\Lambda$ systems in leaky cavities \cite{beige00a,beige00b,tregenna02a} and show how both lead to the similar scaling.

The paper is divided as follows: in section \ref{sec1}, we introduce the set-up where we implement our proposal and establish the general formalism that we use to characterize the operations. In section \ref{sec3}, we describe the logical qubits and computational subspace and show how to build a set of universal quantum gates in the ideal case, that is, without considering decay into other non-guided modes or deviations from Quantum Zeno dynamics \cite{zanardi97a,lidar98a,facchi02a}. Then, in sections \ref{sec4} and \ref{sec5}, we analyze possible error sources, both analytically and numerically, including spontaneous emission and imperfect addressing for the different gates of our proposal. Finally, in section \ref{sec6}, we compare the scaling with the proposal of three-level atoms in leaky cavities already explored in the literature \cite{beige00a,beige00b,tregenna02a}.

\section{General set-up and formalism} \label{sec1}

\subsection{Set-up: waveguide QED}

\begin{figure}[t]
\centering
\includegraphics[width=0.98\textwidth]{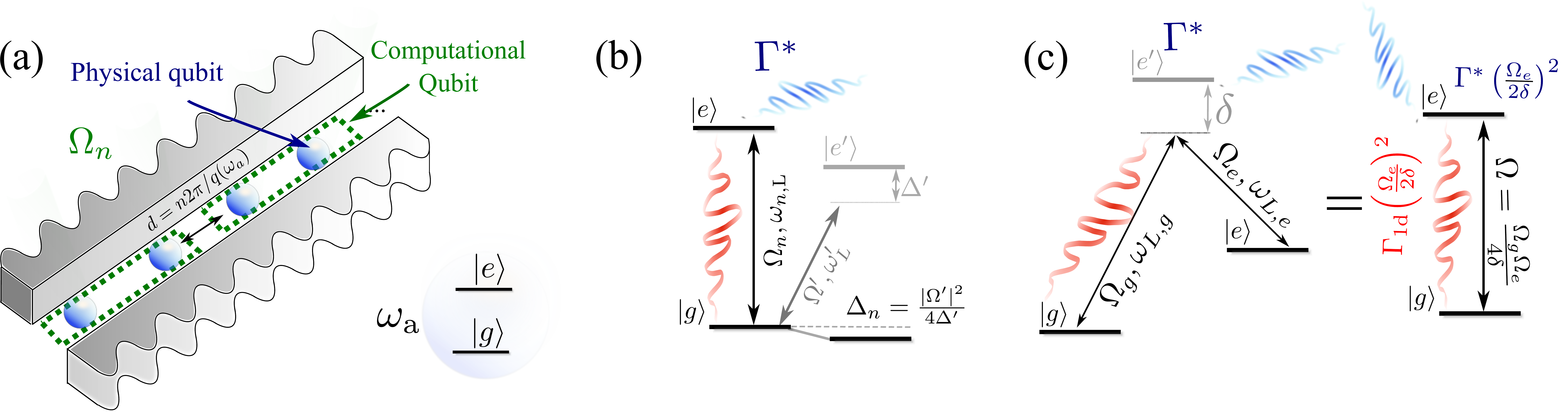}
\caption{(a) TLS (in blue) trapped along a one-dimensional waveguide, addressed by single-site resolved control fields. In green, we depict the pairing that we will use to engineer the computational qubits.(b) Level structure of a TLS with a coherent driving with amplitude (detuning) $\Omega_n$ ($\Delta_n$)  with an additional level to which transitions can be driven off-resonantly to engineer the $\Delta_n \sigma_{ee}^n$ term of the hamiltonian. (c) A three-level system in which the excited state is driven far off-resonantly can be made approximately equivalent to a TLS with modified parameters as shown in the legend.}
\label{fig1}
\end{figure}

The general set-up that we consider is depicted in figure \ref{fig1}a; namely $N$ TLSs, $\{ |g\rangle_n$, $| e \rangle_n \}_{n = 1 \dots N}$, placed at positions $z_n$ and coupled to a 1d field with bosonic annihilation operators $a_q$.  Due to the variety of implementations available nowadays, we will keep the discussion as general as possible without making further assumptions on the nature of the TLS and/or 1d waveguides.

The composite system is described by the Hamiltonian $H = H_0 + H_\II$, where $H_0$ is the free term given by $H_0 = H_\mathrm{qb} + H_\mathrm{field}$,  (using $\hbar = 1$)
\begin{equation}
	H_\mathrm{qb} = \omega_\aa \sum_{n=1}^N \sigma^{n}_{ee}, \
	H_\mathrm{field} = \sum_q \omega_q a^\dagger_q a_q,
	\label{H0}
\end{equation}
where $\omega_\aa$ is the TLS energy, $\sigma_{ij}^n=\ket{i}_n\bra{j}_n$ are atomic operators, and $\omega_q$ is the energy dispersion relation of the waveguide modes. We consider a dipolar coupling of the form
\begin{equation}
	H_\II  = \sum_{n} \left( \sigma_{ge}^n E(z_n) + \mathrm{H.c.} \right),
	\label{Hint}
\end{equation}
with $E(z) = \sum_q g_q (a_q e^{i q z}+\ud{a_q} e^{-i q z})$, and $g_q$ the single photon coupling constant. 
When the system-reservoir coupling is weak (Born approximation) and the evolution of the 1d-reservoir is much faster that the one of the system (Markov approximation), the evolution of $\rho$, the reduced density matrix for the atom-like system, can be described by a Markovian master equation of the form $d \rho /d t = {\cal L}\left[\rho\right]$ \cite{lehmberg70a,lehmberg70b,chang12a,gardiner_book00a}, with the superoperator
\begin{equation}
	\label{mequation1}
	{\cal L}\left[\rho\right] = 
	\sum_{n,m} \Gamma_{n,m} \left( \sigma_{ge}^n \rho \sigma_{eg}^m - \rho \sigma_{eg}^m \sigma_{ge}^n \right)
	+ \mathrm{H.c.} \,,
\end{equation}
where
\begin{equation}
	\Gamma_{n,m} = \frac{\Gamma_{\oned}}{2} e^{i q(\omega_\aa) |z_n - z_m|}\,,
\end{equation}
where $\Gamma_\oned$, the decay into waveguide modes, that we will assume to be larger than the rate of spontaneous emission into all other modes, $\Gamma^* \ll \Gamma_\oned$ as this is the regime we are interested in. Moreover, as the propagation lengths of the waveguide modes for many implementations are long ($L_\mathrm{p} \gg \lambda_\mathrm{a}$), the atoms can be separated several wavelengths apart and can therefore be individually addressed as depicted in figure \ref{fig1}(a). In particular, we assume to control the TLS state through the Hamiltonian (in the interaction picture with respect to $H_\mathrm{qb}$)
\begin{equation}
	H_\mathrm{las}= \sum_{n} \frac{1}{2} \left( \Omega_n \sigma_{ge}^n  + \mathrm{H.c.} \right) + \Delta_n \sigma_{ee}^n,
	\label{eq:Hlas}
\end{equation}
where $\Omega_n$ is the amplitude of the coherent driving (that we consider to be resonant, i.e., $\omega_L=\omega_\aa$) which controls the number of excitations of the system, and $\Delta_n$ is a phase shift interaction term. The latter can be obtained, e.g., in atomic systems, by adding an off-resonant driving to another excited state $\ket{e'}$, as depicted in figure \ref{fig1}(b), which results in an Stark shift $\Delta_n=|\Omega'|^2/(\omega_\aa-\omega_L')$. In general, the way of implementing $\Omega_n$ and $\Delta_n$ will depend on the particular system. 

For completeness, it is worth mentioning that $\Lambda$ systems can also be mapped to effective TLS  by using an off-resonant Raman transition as depicted in figure \ref{fig1}(c). By adiabatically eliminating the excited state $\ket{e'}$, one can formally project the dynamics to the two metastable states, $\{\ket{g},\ket{e}\}$, and find a similar light-matter hamiltonian as the one of equation \ref{Hint}, with the advantage that the effective TLS defined by  $\{\ket{g},\ket{e}\}$ will be long-lived as they are encoded in metastable states. For example, by switching both $\Omega_g$ and $\Omega_e$ at the same time with detuning $\delta(\gg |\Omega_g|,|\Omega_e|)$ as depicted in figure \ref{fig1}(c), we can implement a coherent driving term with effective $\Omega = \frac{\Omega_g \Omega_e^*}{4 \delta}$. By switching $\delta$ in this case big enough one can neglect spontaneous emission processes as they will be proportional to $\Gamma^* \left(\frac{|\Omega_e|^2+|\Omega_g|^2}{4\delta^2}\right)$. Moreover, if we 
switch only $\Omega_e$ and adiabatically eliminate the photonic modes we also obtain an irreversible transition from $\ket{e}\rightarrow\ket{g}$, but with a renormalization of the decay rates $\Gamma_\oned \rightarrow \Gamma_\oned | \frac{\Omega_e}{2 \delta} |^2$ and $\Gamma^* \rightarrow \Gamma^* | \frac{\Omega_e}{2\delta}|^2$. Hence, the Purcell factor $P_\oned = \Gamma_\oned / \Gamma^*$ is unchanged. In that situation our analysis is an alternative implementation to the one developed in Refs. \cite{beige00a,beige00b,tregenna02a}.

\subsection{Decoherence-Free Subspaces}

In the case of equidistant spacing at positions commensurate with the wavelength of the guided mode, i.e.~$z_n =n 2\pi/q(\omega_\aa)$, the effective interaction induced by the waveguide modes yields a pure Dicke model \cite{dicke54a} decay described by 
\begin{equation}
	{\cal L}_\DD \left[ \rho \right] 
	= \frac{\Gamma_\oned}{2} \left(S_{ge} \rho S_{eg} -  S_{eg} S_{ge} \rho \right) + \mathrm{H.c.},
	\label{Dicke}
\end{equation}
where we have introduced the collective spin operator $S_{ge} = \sum_{n=1}^N \sigma^n_{ge}$. The states satisfying $S_{ge} \ket{\Psi}=0$ are \textit{decoherence-free} with respect to the collective dissipation $\mathcal{L}_\DD$. These states can be easily described in the collective spin basis $\{| J, m_J, \alpha_J \rangle \}$, that is the eigenstates of the collective operators $S^2 =\sum_{i={x,y.z}}S_{i}^2$ and $S_z$ with
\numparts
\begin{eqnarray}
	S^2\ket{J,m_J,\alpha_J}&=J(J+1)\ket{J,m_J,\alpha_J}\,,\\
	S_z\ket{J,m_J,\alpha_J}&=m_J \ket{J,m_J,\alpha_J}\,,
\end{eqnarray}
\endnumparts
where $J=N/2,N/2-1,\dots$, $m_J=-J, -J+1,\dots,J$. The index $\alpha_J$ is introduced because the states in the collective spin basis are degenerate, with degeneracy  given by: $\alpha_J = 1,\dots,  \binom{N}{J}-\binom{N}{J-1}$. It is easy to observe in this basis that the states $\ket{J,-J,\alpha_J}$ are decoherence free, and therefore span the \textit{decoherence-free subspace} (DFS).

The DFS has a dimension of $\binom{N}{N/2}$ (assuming even atom number $N$), and is composed of all the possible states which are antisymmetric with the permutation of two atoms. Thus, an alternative way of characterizing the DFS is to consider all possible (tensor products) of singlet states
\begin{equation}
	\ket{A_{m,n}} = \left(\ket{e}_m \otimes \ket{g}_n-\ket{g}_m \otimes \ket{e}_n\right)/\sqrt{2}\,,
\end{equation}
where $m,n$ denote the atomic positions of the pair of atoms that form the singlet. This characterization makes it more difficult to describe an orthonormal basis of the DFS. However, we show in the next section that it is convenient to define our computational subspace.

\subsection{Quantum Zeno dynamics using Liouvillian perturbation theory} \label{sec2}

We are interested in the regime where the collective dissipation induced by  $\mathcal{L}_\mathrm{D}$, with characteristic timescale $\Gamma_{\oned}^{-1}$, dominates over any possible perturbation of the system, $\mathcal{L}_\mathrm{pert}$, with characteristic timescale $\tau\gg\Gamma_{\mathrm{1d}}^{-1}$. Under these assumptions, any state outside of the DFS will only be virtually populated due to the strong dissipation and therefore the dynamics will be restricted to the slow subspace, i.e., the DFS. Mathematically, we formalize this intuitive picture by defining a projection superoperator $\mathbb{P}$ (with $\mathbb{P}^2 = \mathbb{P}$) satisfying: $\mathbb{P}{\cal L}_\DD={\cal L}_\DD \mathbb{P}=0$ that projects out the fast dynamics yielding only the effective evolution in the slow subspace. It is then possible to integrate out the fast dynamics (see Appendix A) arriving to an effective master equation given by
\begin{equation}
	\frac{\partial \PP\rho}{\partial t}=\mathcal{L}_\mathrm{eff} [\PP\rho]= \Big(\PP  \mathcal{L}_\mathrm{pert} \PP +\PP  \mathcal{L}_\mathrm{pert} \QQ \frac{1}{{-\cal L}_\DD}\QQ  \mathcal{L}_\mathrm{pert} \PP +\mathrm{O}\left(\tau^{-3}/\Gamma_\oned^2 \right)\Big) \rho\,.\label{eq:effemastereq}
\end{equation}

This result to first order (left hand term in the brackets) accounts for the ideal \emph{Quantum Zeno dynamics}  \cite{zanardi97a,lidar98a,facchi02a}. The second order in perturbation theory then yields correction terms mainly coming from slightly populating the (super)radiant states.

In our case, there will be two types of perturbations, namely,
\begin{itemize}
	 \item The Hamiltonian $\mathcal{L}_\mathrm{pert}[\cdot]= - \rmi \left[H_{\mathrm{las}}, \cdot \right]$ to control the atomic state. This results to first order in an effective Hamiltonian $H_{\mathrm{eff}}=\mathcal{P} H_{\mathrm{las}} \mathcal{P}$ that couples only atomic states within the DFS. Here, we introduced the projection onto the DFS for pure states $\mathcal{P} = \sum_i \ket{\mathrm{d}_i}\bra{\mathrm{d}_i}$, where the states $\ket{\mathrm{d}_i}$ form an orthonormal basis of the DFS. We use this effective laser coupling to control the atomic state of the ensemble. Besides, there is a second order correction resulting from $H_{\mathrm{las}}$ that will be relevant for the analysis of the error probability of our proposal as we show in section \ref{sec4}.
 
	 \item The contribution of the emission of photons to other radiative modes different from the guided mode of the waveguide that we embed into a single decay rate, $\Gamma^*$ and describe through the Liouvillian 
	 \begin{equation}
		 \mathcal{L}_\mathrm{pert}[\rho]=\mathcal{L}_{*}[\rho]=\sum_n \frac{\Gamma^*}{2}(\sigma_{ge}^n\rho\sigma_{eg}^n-\rho \sigma_{ee}^n+\mathrm{H.c.}).
		 \label{eq:SpEm}
	 \end{equation}
 This contribution is relevant for the error analysis of the gates in section \ref{sec4}.
\end{itemize}

\section{Universal Single- and Two-Qubit Gates} \label{sec3}

In this Section, we show how to engineer a set of universal gates, i.e.,  defined by any arbitrary single-qubit rotation and a controlled gate  \cite{nielsen_book00a}, using the effective evolution $H_{\mathrm{eff}}$ within the DFS that appear in our waveguide QED setup. Firstly, due to the large degeneracy of the DFS, we need to define a set of logical qubits that will expand our computational subspace. Then, we show how to choose $\{\Omega_n,\Delta_n\}$ such that they define a set of universal one and two-qubit gates, namely, the phase and Pauli-$X$ (and $Y$) gate and the controlled-$(-Z)$ gate.  A summary of the parameters for these gates can be found in table~\ref{tab:GateSum}. For completeness, we also give the parameters for other gates such as the Hadamard or SWAP gates. The former can be easily constructed because all single qubit rotations can be performed and the latter is constructed through the same idea as the controlled-$(-Z)$ gate. Due to the degeneracy of the DFS, the challenge lies in defining 
operations within the computational subspace, without populating the rest of the states within the DFS.  In section \ref{sec4}, we revisit the problem and consider the effect of spontaneous emission and second order corrections to the Zeno dynamics that ultimately limit the fidelity of the operations.

\begin{table}[!b]
	\caption{Summary of the optimal parameter settings for the $\pi/8$-gate $T$, the Pauli gates $X, \ Y,\ Z$, the Hadamard gate $H$, the SWAP-gate and the controlled-$(-Z)$ gate. The subindex denotes on which logical qubit the gate acts. The settings for the Rabi couplings and detunings are denoted by $x_{ij}^\pm = \frac{1}{2}\left(x_i \pm x_j\right)$. $T$ is the duration for which the operation is applied to obtain the corresponding gate. In general $\Delta_\mathrm{D}$ is a large detuning (see equation \ref{eq:DeltaD}), that prevents transitions to other states and $\Omega_n= 0$ for $n \geq 3 $.}\label{tab:GateSum}
	\begin{indented}
		\item[]
		\begin{tabular}{l|cccccc|c|c}
			\br
			Gate & $\Omega_{12}^-$ & $\Omega_{12}^+$ & $\Delta_{12}^+$ & $\Delta_{12}^-$ & $\Delta_{34}^+$ & $\Delta_{34}^-$ & $T$ & $\Delta_{n\geq5}$\\
			\br
			$ T_1$ & 0 & 0 & -$\Delta_\mathrm{T}$ & 0 & 0 & 0 & $\frac{\pi}{4 \Delta_\mathrm{T}}$& 0 \\
			$ Z_1$ & 0 & 0 & $\Delta_\mathrm{Z}$ & 0 & 0 & 0 & $\frac{\pi}{\Delta_Z}$ & 0 \\
			$ X_1$ & $\Omega_\mathrm{X} \in \mathbb{R}$ & 0 & 0 & 0 & $\Delta_\mathrm{D}$ & 0 & $\frac{\pi}{\sqrt{2} \Omega_\mathrm{X}}$ &  $\Delta_\mathrm{D}$ \\
			$ Y_1$ & $\Omega_\mathrm{Y} \in \rmi \mathbb{R}$ & 0 & 0 & 0 & $\Delta_\mathrm{D}$ & 0 & $\frac{\pi}{\sqrt{2} |\Omega_\mathrm{Y}|}$ &  $\Delta_\mathrm{D}$ \\
			$ \propto H_1$ & -$\Omega_\mathrm{H} \in \mathbb{R}$ & 0 & $\sqrt{2} \Omega_\mathrm{H}$ & 0 & $\Delta_\mathrm{D}$ & 0 & $\frac{\pi}{2 \Omega_\mathrm{H}}$ &  $\Delta_\mathrm{D}$ \\
			SWAP$_{12}$ & 0 & 0 & 0 &  $\Delta_\mathrm{S}$ & 0 &  $\Delta_\mathrm{S}$ & $\frac{\pi}{\Delta_\mathrm{S}}$ &  $\Delta_\mathrm{D}$ \\
			C($-Z$)$_{12}$ & 0 & 0 & 0 & $\Delta_\mathrm{C}$ & 0 & 0 & $\frac{\pi 2 \sqrt{2}}{\Delta_\mathrm{C}}$ & $\Delta_\mathrm{D}$ \\
			\br
		\end{tabular}
	\end{indented}
\end{table}

\subsection{Definition of Logical Qubits and Computational Subspace}

Before finding the appropriate gates, we first need to define the logical qubits within the DFS. In principle, assuming an even number of atoms $N$, the dimension of the DFS is $\binom{N}{N/2}$ and therefore, it is possible to encode $\log_2 \binom{N}{N/2}$ logical qubits. However, it is more useful to restrict the computational subspace to a smaller set of states in order to achieve universal quantum computation. As the DFS is spanned by all (tensor) products of singlet states over two atoms, it is natural to define the logical qubits as
\numparts
\begin{eqnarray}
	\ket{0}_{j}^{\mathrm{L}} \equiv & \ket{G_{j,j+1}} = \ket{g}_j \otimes \ket{g}_{j+1} \label{eq:logicalQB1} \\
	\ket{1}_{j}^{\mathrm{L}} \equiv & \ket{A_{j,j+1}} =  \left(\ket{e}_j \otimes \ket{g}_{j+1}-\ket{g}_j \otimes \ket{e}_{j+1}\right)/\sqrt{2}.
	\label{eq:logicalQB2}
\end{eqnarray}
\endnumparts

It is instructive to consider particular examples to see how the DFS and computational space look, i.e., for the case of $N=2$ and $N=4$ atoms.

\begin{figure}
	\centering
	\includegraphics[width=0.98\textwidth]{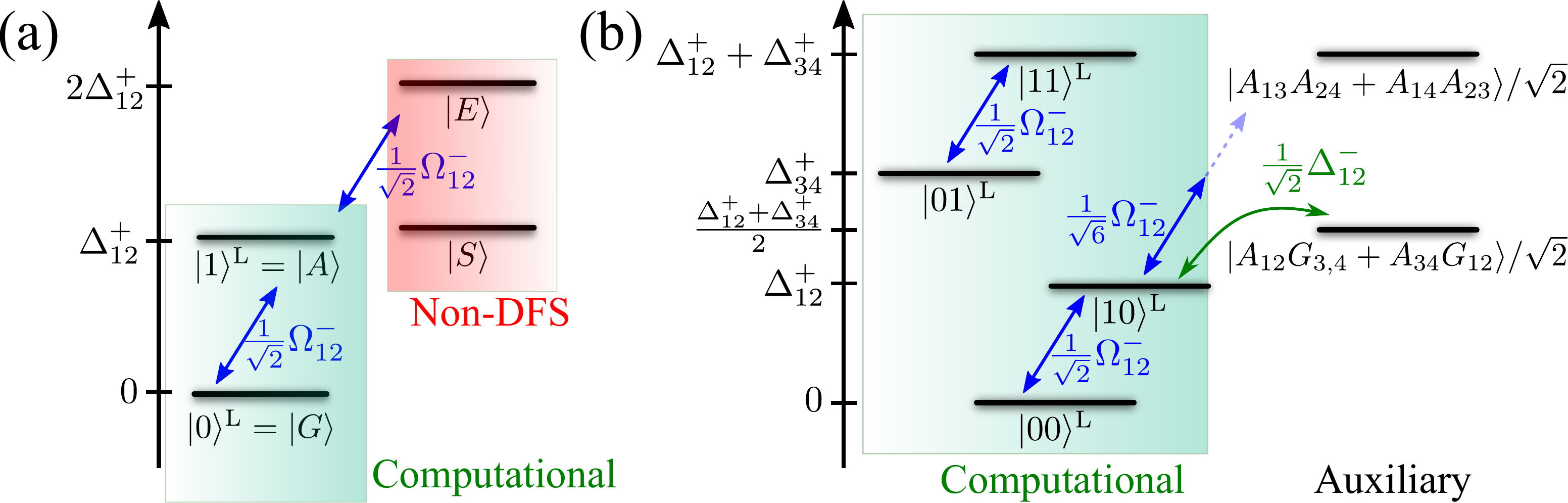}
	\caption{(a) Hilbert space of two TLS in the basis of non-DFS and DFS states with energies in the interaction picture with respect to $H_\mathrm{qb}$. The blue arrows denote the transitions necessary for the Pauli-X gate. The triplet states $\ket{E} = \ket{ee}$ and $\ket{S} = \ket{eg+ge}/\sqrt{2}$ are not inside the DFS. (b) DFS of 4 TLS consists of 6 states that split into the $2^2$-dimensional computational subspace and two states in the auxiliary subspace. The blue [green] arrows denote the transitions necessary for the Pauli-X [and Controlled-(-Z)] gate.} \label{fig2}
\end{figure}

\paragraph{\textbf{Two atoms:} } In this case it is easy to plot the complete Hilbert space (including states outside DFS) as it consists only of 4 states as depicted in figure \ref{fig2}(a). The separation into DFS states and non-DFS states is easily done in the familiar singlet-triplet basis. The DFS consists of two states: the one with two atoms in the ground state and the singlet state, i.e., the antisymmetric combination of one single excited state. Thus, we can encode one logical decoherence free qubit. The other two states are superradiant, i.e., they decay with an enhanced decay rate of $2 \Gamma_\oned$.

\paragraph{\textbf{Four atoms:} } The complete Hilbert space consists of $2^4=16$ states, and the dimension of the DFS, shown in figure \ref{fig2}(b), is $\binom{4}{2} = 6$.  As aforementioned, we want to use as computational subspace the tensor product of the antisymmetric pairs described in equation \ref{eq:logicalQB1} and \ref{eq:logicalQB2} , which consist only of $2^{2}=4$ states. This is why in this situation we have to distinguish within the DFS between the computational space and the additional states  that must be either decoupled or used as auxiliary states. In particular, for $N=4$ the auxiliary states are $\ket{A_{1,2}G_{3,4}+A_{3,4}G_{1,2}}/\sqrt{3}$ and  $\ket{A_{1,3}A_{2,4}+A_{1,4}A_{2,3}}/\sqrt{2}$.

\paragraph{\textbf{For $N>4$ atoms:} }
In general (for $N>2$) the dimension of the DFS, $\binom{N}{N/2}$ (for even $N$), is larger than the one of the computational subspace, $2^{N/2}$. Thus, one can split the projection onto the DFS, $\mathcal{P}$, into two orthogonal projections, i.e. one into the computational subspace $\mathcal{P}_\mathrm{CS}$ and its orthogonal counterpart $\mathcal{Q}_\mathrm{CS}$:
\numparts
\begin{eqnarray}
	\mathcal{P} =& \mathcal{P}_\mathrm{CS} + \mathcal{Q}_\mathrm{CS},\\
	\mathcal{P}_\mathrm{CS} =& \sum_{j\ \mathrm{odd}} \ket{0}^\mathrm{L}_j \bra{0} + \ket{1}^\mathrm{L}_j \bra{1}\,,
\end{eqnarray}
\endnumparts
such that the effective Hamiltonian can be written as follows
\begin{equation}
	H_\mathrm{eff} = 
	  \mathcal{P}_\mathrm{CS} H_{\mathrm{las}} \mathcal{P}_\mathrm{CS} 
	+ \left( \mathcal{P}_\mathrm{CS} H_{\mathrm{las}} 	\mathcal{Q}_\mathrm{CS} + \mathrm{H.c.} \right) 
	+ \mathcal{Q}_\mathrm{CS} H_{\mathrm{las}} 	\mathcal{Q}_\mathrm{CS},
\end{equation}
which separates the transitions within the computational [auxiliary] subspace $\mathcal{P}_\mathrm{CS} H \mathcal{P}_\mathrm{CS}$ [$\mathcal{Q}_\mathrm{CS} H 	\mathcal{Q}_\mathrm{CS}$] and the coupling between these two subspaces: $\mathcal{P}_\mathrm{CS} H\mathcal{Q}_\mathrm{CS}$.
This separation will be useful to argue that we can make operations within the DFS even in the situations with $N>4$ as we will show afterwards. For a general situation, it is easy to show that by projecting $H_{\mathrm{las}}$ into the computational subspace we obtain an effective evolution inside the computational subspace given by:
\begin{equation}
	\mathcal{P}_\mathrm{CS} H_{\mathrm{las}}\mathcal{P}_\mathrm{CS}
	= \sum_{j \ \mathrm{odd}} 
		\left(\frac{\Omega_{j,j+1}^{-}}{\sqrt{2}} \ket{1}^\mathrm{L}_j\bra{0} + \mathrm{h.c.} \right)
		+ \Delta_{j,j+1}^+ \ket{1}^\mathrm{L}_j \bra{1},
\end{equation}
where we used the following notation $x_{i,j}^\pm = \frac{1}{2} \left(x_i \pm x_{j}\right)$ to abbreviate the combination of parameters. 

\subsection{Single-Qubit Gates}

The goal is to find the $\{\Omega_n,\Delta_n\}$ such that they define both the phase and Pauli-$X$ (and $Y$) gates over the computational subspace. 

\paragraph{\textbf{Two atoms:} }
This is the simplest situation because the size of the computational space is the same as the one of the DFS. In this case (see also \cite{beige00b}),  a phase shift gate on the logical qubit ($\alpha \ket{0}^\mathrm{L} + \beta \ket{1}^\mathrm{L}\rightarrow \alpha \ket{0}^\mathrm{L} + \beta \mathrm{e}^{-\rmi \phi} \ket{1}^\mathrm{L}$) is obtained by applying $\Omega_{12}^-= 0$,  $\Delta_{12}^+ \neq 0$ for a time $T = \frac{\phi}{\Delta_{12}^+}$. Pauli-$X$ rotations (plus a phase) are obtained for $0 \neq \Omega_{12}^- \in \mathbb{R}$, $\Delta_{12}^+ = 0$ and time $T = \frac{\pi}{\sqrt{2} \Omega_{12}^-}$.  Note that to avoid errors in both cases, one should also set $\Omega_{12}^+ = \Delta_{12}^- = 0$ as will be discussed in section~\ref{sec4}. The Pauli $Y$ can be obtained as the $X$ just by using $i\Omega_{12}^- \in\ \mathbb{R}$, so that we will not discuss it further.

\paragraph{\textbf{Four atoms:} }
In this case, the way to do phase gates and rotations is the same as for the two atom case. However, in the case of the rotations,  states within the computational subspace couple to auxiliary states for more than two atoms. In particular, the state $\ket{10}^\mathrm{L}$ is coupled to the auxiliary state $\ket{A_{1,3}A_{2,4}+A_{1,4}A_{2,3}}/\sqrt{2}$ for $\Omega_{12}^- \neq 0$ as shown in figure \ref{fig2}.
However, this transition can be made far off-resonance by setting $|\Delta_{34}^+| \gg |\Omega_{12}^-|$. This results in an additional error rate $\frac{|\Omega_{12}^-|^2}{2 \Delta_{34}^+}$ that will be considered when calculating the fidelity of the operation.\footnote{In fact, this argument can be reversed to excite the auxiliary state from the computational state $\ket{10}^\mathrm{L}$ with the choice $\Delta_{34}^+ = 0$ and $\Delta_{12}^+ \gg \Omega_{12}^-$.}

\paragraph{\textbf{For $N>4$ atoms:} }
Again in the case of rotations, transitions to states outside the computational subspace in the ideal case ($\Omega_{12}^+ = \Delta_{12}^- = 0$) are possible when $\Omega_{12}^- \neq 0$, that is when
\begin{equation}
	\mathcal{Q}_\mathrm{CS} H \mathcal{P}_\mathrm{CS}
	=\mathcal{Q}_\mathrm{CS} H_\mathrm{eff} \mathcal{P}_\mathrm{CS} \neq 0\,,
\end{equation}
where we use that $\mathcal{P}_\mathrm{CS} \mathcal{P} = \mathcal{P}_\mathrm{CS}$.
However, the transitions to these states can be made far off-resonant by setting
\begin{equation}
	\Omega_n = 0, \ \mathrm{and} \  |\Omega_{12}^-| \ll \Delta_n 
	= \Delta_\mathrm{D} \ll \Gamma_\oned, \ n \geq 3,
\end{equation}
because the auxiliary states inside the DFS that the computational subspace couples to extend over more than two atoms \footnote{The auxiliary states necessarily extend over more than two atoms, because it is orthogonal to the logical qubits and therefore contains excited (superradiant) triplet states in the ``pairing'' of the atoms. An antisymmetric combination of such states can be in the DFS, but not in the computational subspace, and necessarily extends over multiple atom ``pairs''.} and can therefore be detuned as
\begin{equation}
	\mathcal{Q}_\mathrm{CS} H \mathcal{Q}_\mathrm{CS} \sim \Delta_D \mathcal{Q}_\mathrm{CS}\,,
	\label{eq:DeltaD}
\end{equation}
while keeping the desired transition driven by $\Omega_{12}^-$ as resonant. One has to make sure that the Stark-shift introduced by this off-resonant transition is small and possibly correct the detuning that it will induce by choosing appropriately the applied laser frequency $\omega_L$.

\subsection{Controlled Pauli-Z}
For universal quantum computation, a controlled two-qubit gate is required. In this case, the minimal system to encode the operation is the $N=4$ atom case, where two decoherence-free logical qubits can be obtained.

\paragraph{\textbf{Four atoms:} }  
In order to build the controlled-$Z$ gate, we use one of the auxiliary states, $\ket{A_{1,2}G_{3,4}+A_{3,4}G_{1,2}}/\sqrt{2}$. Now, it is possible to drive only the transition between this state and $\ket{10}^\mathrm{L}$ without affecting the other states within the DFS by the choice $\Omega_n = 0$, $\Delta_3=\Delta_4 = 0$ and $\Delta_1 = -\Delta_2 \neq 0$. A $\pi$-pulse on the state $\ket{10}^\mathrm{L}$ leads to a relative phase of $-1$ on this state, i.e.
\begin{equation}
	\ket{10}^\mathrm{L}_{1,3} \longrightarrow -\ket{10}^\mathrm{L}_{1,3},
\end{equation}
for $\frac{1}{\sqrt{2}} \Delta_{12}^- T = 2 \pi$ without affecting the other states of the computational subspace. Hence, we have defined a a controlled controlled-$(-Z)$ gate which is equivalent up to single qubit unitaries to a CNOT-gate \cite{nielsen_book00a}.

\paragraph{\textbf{For $N>4$ atoms:} }
One can restrict the dynamics to the subspace of four atoms in a similar way as for the single-qubit rotations. With the choice of 
\begin{equation}
	\Omega_n = 0, \ \mathrm{and} \ |\Delta_{12}^-| \ll \Delta_n =\Delta_\mathrm{D} \ll \Gamma_\oned , \ n\geq 5,
\end{equation}
transitions to states over more than four atoms are far off-resonant.
As before, this adds an error rate proportional to $\frac{|\Delta_{12}^- |^2}{\Delta_\mathrm{D}}$ with a proportionality factor depending on the coupling strength after the projection onto the DFS.

\section{Error analysis: spontaneous emission and imperfect addressing.} \label{sec4}

So far we have considered only the interaction within the ideal Quantum Zeno Dynamics, where the only possible sources of error were due to the larger dimension of the DFS with respect to the computational space. In this section, we take into account other sources of errors that will be present in most of the implementations, namely, i) errors coming from spontaneous emission to other modes, with rate $\Gamma^*$, included through ${\cal L}_{*}\left[\rho\right] $ as in equation~\ref{eq:SpEm}; ii) errors from deviations from the Zeno Hamiltonian, attributed to photons emitted to the waveguide from the small population present in the states outside the DFS; iii) errors that may arise from an imperfect control of the laser parameters $\{\Omega_n,\Delta_n\}$. In what follows, we assume to work in a regime with $P_{\oned}\gg 1$, such that the following parameter hierarchy can be satisfied: $\Gamma^*\ll ||H_\mathrm{eff}|| \ll \Gamma_\oned$.

This section discusses, for each gate separately, the numerical results and their analytical approximation of the fidelity between the theoretical final (goal) state, $\ket{\psi_f}$, and the real atomic state, $\rho$, obtained after the gate operation, i.e., $F = \langle \psi_\mathrm{f} | \rho | \psi_\mathrm{f} \rangle^{1/2}$. 
The numerical results are obtained by solving the master equation in second order perturbation theory (see equation \ref{eq:effemastereq}). We have checked numerically that this is a good approximation in the parameter ranges considered throughout this manuscript. To obtain the analytical approximations, we used the effective non-hermitian Hamiltonian that can be obtained from the second order master equation (see appendix A).

\subsection{Phase Shift Gate}

\begin{figure}[t]
	\centering
	\includegraphics[width=0.98\textwidth]{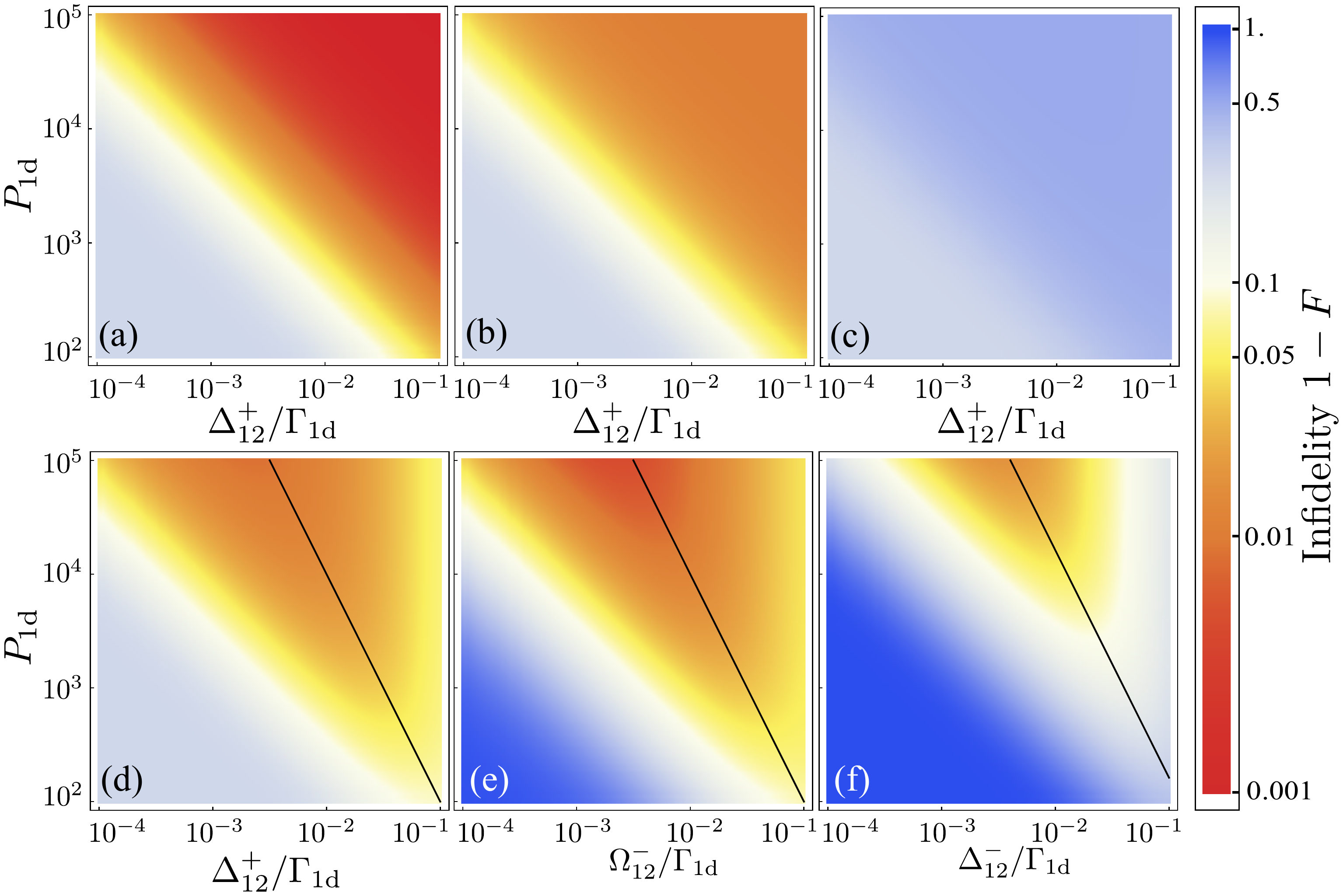}
	\caption{Infidelities of single and two qubit gates for $N=4$ atoms. (a)-(d) Infidelity of a  $\pi/2$ phase shift on the first logical qubit of four atoms on the state $\ket{10+00}^\mathrm{L}/\sqrt{2}$ for (a) $\Delta_{12}^- = 0$ and $\Delta_{34}^+ = 0$, (b) $\Delta_{12}^- = 0.1 \Delta_{12}^+$ and $\Delta_{34}^+ = 0$, (c) $\Delta_{12}^- = \Delta_{12}^+$ and $\Delta_{34}^+ = 0$, (d) $\Delta_{12}^- = \Delta_{12}^+$ and $\Delta_{34}^+ = 20 \Delta_{12}^+$. (e) Infidelity of a single qubit Pauli-X gate on the first qubit of four atoms on the state $\ket{00}^\mathrm{L}$ for  $\Delta_{34}^+ = 10 \Omega_{12}^-$. (f) Infidelity of the controlled-$(-Z)$ gate for four atoms (2 qubits) acting on the state $\ket{10+11}^\mathrm{L}/\sqrt{2}$. The black lines in (d)-(f) represent the scaling of the coupling strength $\Delta_{12}^-$, $\Omega_{12}^-$ and $\Delta_{12}^-$, respectively, for the minimal infidelity with $P_\oned^{-1/2}$.} \label{fig3}
\end{figure}

For the phase shift gate we must set $\Omega_n = 0$ for all $n$ and $\Delta_n = \Delta_\mathrm{D} \gg \Delta_{12}^-$ for all $n \geq 3$ to avoid errors from transitions to auxiliary states. By choosing $\Delta_1 = \Delta_2$, i.e., $\Delta_{12}^-=0$, no errors (from second order perturbation theory) occur because the computational states do not couple to the radiant ones. However, it is instructive to consider the errors that appear for situations where $\Delta_{12}^-\neq 0$, e.g, because of imperfect addressing, as this yields a useful understanding about how to deal with situations where the second order correction cannot be avoided.

\paragraph{\textbf{Two atoms:}}
For the simplest situation the additional errors due to imperfect addressing, i.e., $\Delta_{12}^-\neq 0$, enter at a rate proportional to $|\Delta_{12}^-|^2/\Gamma_\oned$ through the same error channel as the spontaneous emission into all other modes with rate $\Gamma^*$, that is, via the quantum jump operator $\ket{0}^L \bra{1}$. Then, the infidelity, i.e., $1-F$, for a $\pi/2$ phase shift of the first logical qubit on the normalized state $\alpha \ket{0}^\mathrm{L} + \beta \ket{1}^\mathrm{L}$ can be approximated by
\begin{equation}
	1-F
	\approx \frac{|\beta|^2}{4} \frac{\pi}{\Delta_{12}^+} 
		\left( \Gamma^* + 4 \frac{| \Delta_{12}^-|^2}{\Gamma_{1\mathrm{D}}} \right).
	\label{eq:FidAppPh}
\end{equation}
One observes, that in the ideal case, $\Delta_{12}^- = 0$, the infidelity can be arbitrarily close to $0$ for large $\Delta_{12}^+$. If $\Delta_{12}^-$ is not negligible, the transition strength $\Delta_{12}^+$ cannot be chosen arbitrarily large to decrease the infidelity. For example, in the worst case scenario where $\Delta_{12}^-  = \Delta_{12}^+$ this results in an optimal infidelity scaling $\frac{|\beta|^2 \pi }{2} P_\oned^{-1/2}$ for $\Delta_{12}^+ = \Delta_{12}^-=\frac{1}{2} \sqrt{\Gamma^* \Gamma_\oned}$.

\paragraph{\textbf{Four atoms:} }  
A similar behaviour can be obtained by choosing $\Delta_{12}^- = 0$ such that the infidelity is arbitrarily close to $0$ (see figure \ref{fig3}a). Slight deviations from this ideal value do not change this behaviour drastically (see figure \ref{fig3}b). However, when $\Delta_{12}^-$ is not negligible, it leads to two types of errors that decrease the fidelity (see figure \ref{fig3}c): i) virtual population of non-DFS states, which leads to an error rate proportional to $|\Delta_{12}^-|^2/\Gamma_\oned$ as for two atoms; and ii) transitions to auxiliary states, in particular $\ket{A_{1,2}G_{3,4}+A_{3,4}G_{1,2}}/\sqrt{2}$. The latter can be made far off-resonance by applying a detuning on the second qubit such that $|\Delta_{12}^-| \ll |\Delta_{34}^+| \ll \Gamma_\oned$. With a large off-resonance ratio $r_\Delta = |\Delta_{34}^+ / \Delta_{12}^-| \gg 1$, one still achieves a small infidelity (see figure \ref{fig3}d).

\begin{figure}[t]
	\centering
	\includegraphics[width=0.98\textwidth]{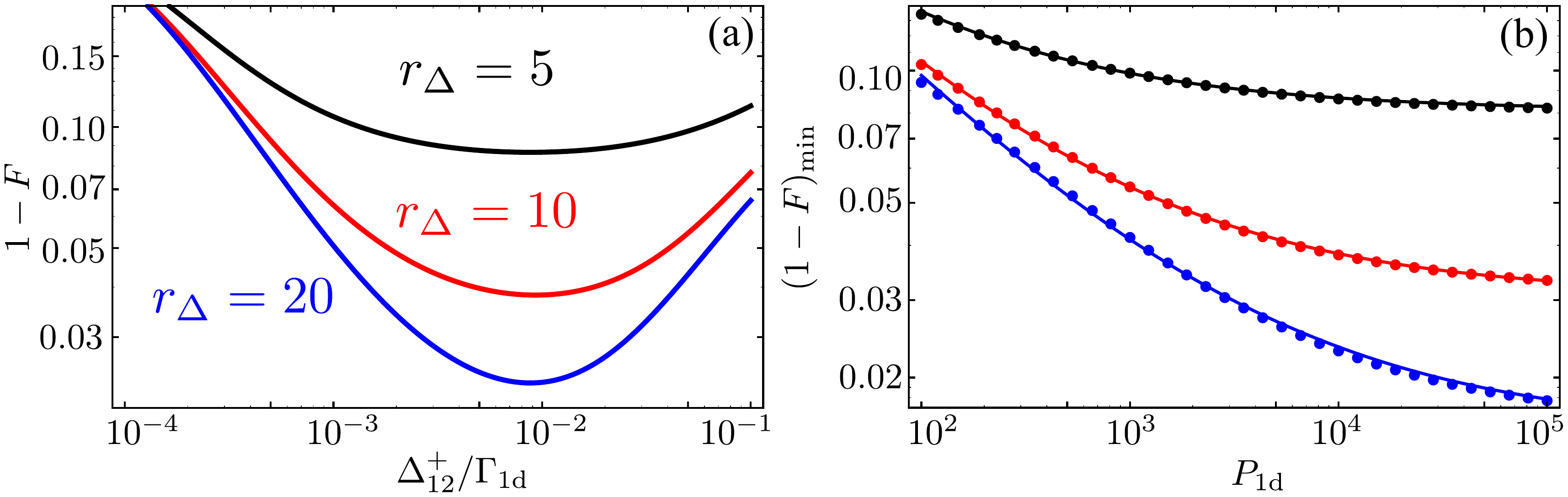}
	\caption{Dependence of infidelity on off-resonance ratio $r_\Delta=|\Delta_{34}^+ / \Delta_{12}^-|$ for $\pi/2$ phase shift on first qubit with $\Delta_{12}^- = \Delta_{12}^+$. (a) Infidelity varied over the coupling strength $\Delta_{12}^+$ for $P_\oned=10^4$.  The results correspond to $r_\Delta=5$ (black), $r_\Delta=10$ (red) and $r_\Delta=15$ (blue). (b) Minimal infidelity depending on the Purcell Factor $P_\oned$. The numerical results (points) corresponding to the values as in (a) fit well with the approximation (line) of $(1-F)_\mathrm{min} \propto  P_\oned^{-1/2}+ C(r_\Delta)$, where the $C(r_\Delta)$ is a number which does not depend on $P_\oned$.}\label{figSM1}
\end{figure}

As shown in figures~\ref{fig3}a-d, the detuning of the third and fourth atom is important when $\Delta_{12}^-$ cannot be neglected. As expected, the infidelity decreases by increasing the off-resonance ratio $r_\Delta$ (see figure~\ref{figSM1}a).
For large enough $r_{\Delta}$, the infidelity can be analytically approximated by%
\begin{equation}
	1-F
	\stackrel{r_\Delta \rightarrow \infty}{\longrightarrow} \frac{\pi}{8} 
		\left( \frac{\Gamma^*}{|\Delta_{12}^+|} + 2 \frac{|\Delta_{12}^+|}{\Gamma_\oned} \right).
\end{equation}
This leads to a minimal infidelity $\propto P_\oned^{-1/2}$ (see figure~\ref{figSM1}b) for $\Delta_{12}^+ = \Delta_{12}^- = \sqrt{\Gamma^* \Gamma_\oned/2}$.

\subsection{Pauli-X Gate}
For rotations around the $x$-axis, we set $\Delta_{1} = \Delta_2= 0$, and $\Delta_n = \Delta_\mathrm{D} \gg \Omega_{12}^-$ for all $n\geq 3$ to avoid transitions to auxiliary states. In contrast to the phase shift gate, even in the ideal case, $\Omega_{12}^+ = 0$, errors will occur because  $\Omega_{12}^-$ couples to state outside the DFS, as shown schematically in figure \ref{fig2}(a) for the two atom case. Moreover, we also include a short discussion on deviations due to imperfect control on $\Delta_{1(2)} \neq 0$ and $\Omega_{12}^+$.

\paragraph{\textbf{Two atoms:} }  
Using $\Omega_{12}^+  = 0$, the error rate from deviations from the Zeno Hamiltonian enters in the same way as from the spontaneous emission into all other modes, that is via the quantum jump operator $\ket{0}_\mathrm{L}\bra{1}$. The corresponding decay rate is $( |\Delta_{12}^-|^2 + |\Omega_{12}^-|^2 / 2) /\Gamma_\oned$. The error from  $\Omega_{12}^+  \neq 0$ enters differently, but can still be included in the estimation of the infidelity. Neglecting the errors from $\Delta_{12}^+ \neq 0$, the infidelity for a Pauli-$X$ gate ($\frac{1}{\sqrt{2}} |\Omega_{12}^-| T = \pi/2$) on state $\ket{1}_\mathrm{L}$ can be approximated by
\begin{equation}
	 1-F
	\approx \frac{1}{2} \frac{\pi}{\sqrt{2} |\Omega_{12}^-|} \left( \Gamma^* + \frac{| \Omega_{12}^- |^2}{2 \Gamma_{1\mathrm{D}}} +  \frac{| \Delta_{12}^- |^2}{\Gamma_{1\mathrm{D}}} + \frac{| \Omega_{12}^+ |^2}{2 \Gamma_{1\mathrm{D}}}\right) \equiv \varepsilon_0 \,.
\end{equation}
In the ideal case of perfect control of addressing parameters, i.e.,  $\Delta_{12}^- = \Omega_{12}^+ = 0$, the minimal value of the infidelity, proportional to $P_\oned^{-1/2}$, is obtained at $|\Omega_{12}^-| = \sqrt{2 \Gamma^* \Gamma_{1\mathrm{D}}}$ as shown in red circles of figure \ref{fig4}. Note, that even for $\Omega_{12}^- = \Omega_{12}^+$ and $\Delta_{12}^- = 0$, the infidelity is still proportional to $P_\oned^{-1/2}$.

\begin{figure}[tb]
	\centering
	\includegraphics[width=0.98\textwidth]{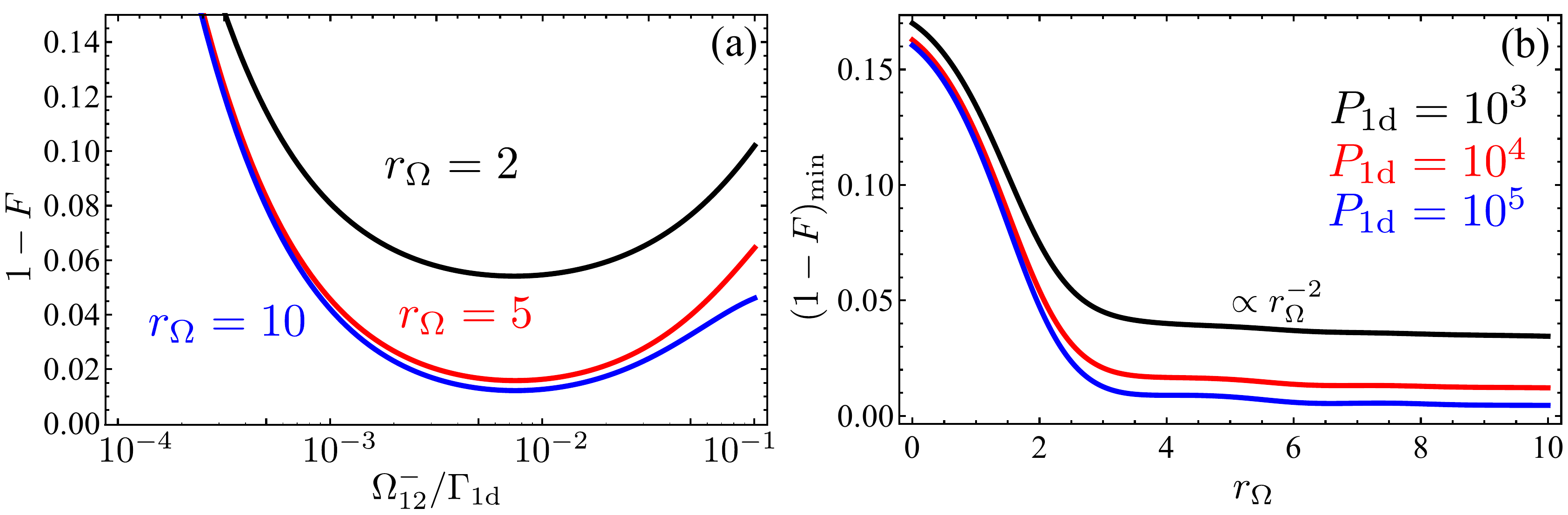}
	\caption{(a) Numerical calculation of infidelity of Pauli-X gate on the first qubit on the state $\ket{00}^\mathrm{L}$ for $N=4$ atoms with $\Delta_{34}^+ = 10 \Omega_{12}^-$ and $P_\oned = 10^4$ for different values of the off-resonance ratio, that is $r_\Omega=2$ (black), $r_\Omega=5$ (red) and $r_\Omega=10$ (blue). (b) Scaling of the minimal infidelity for the same values as in (a) for different Purcell Factors, that is $P_\oned = 10^3$ (black), $P_\oned = 10^4$ (red) and $P_\oned = 10^5$ (blue).} \label{figSM2}
\end{figure}

\paragraph{\textbf{Four atoms:} }  
In this case apart from the transitions out of the DFS, $\Omega_{12}^-$ also couples states inside the DFS, but out of the computational space (see figure \ref{fig2}) such that we need to detune these processes to achieve the rotations. As already explained in the previous section, this can be done by setting $|\Omega_{12}^-| \ll |\Delta_{34}^+| \ll \Gamma_\oned$. As expected, the infidelity decreases when increasing the off-resonance ratio $r_\Omega= |\Delta_{34}^+ / \Omega_{12}^-|$ (see figure~\ref{figSM2}). For large enough ratios $r_\Omega (\gtrsim 4)$, the infidelity can be approximated by 
\begin{equation}
	1-F \approx \varepsilon_0 + \frac{\alpha}{r_\Omega^2},
\end{equation}
where the constant $\alpha = \mathcal{O}(1)$ can be obtained through a numerical fit.
The infidelity of a $\pi/2$-pulse on the state $\ket{00}^\mathrm{L}$ is plotted in figure \ref{fig3}e, whereas the minimal infidelity is shown to scale with $P_\oned^{-1/2}$ in figure \ref{fig4}.

\subsection{Controlled Pauli-Z}

For the controlled-$(-Z)$ gate, we set $\Omega_{n} = 0$ for all $n$, $\Delta_{12}^+ =0 $ and $\Delta_{n} = \Delta_\mathrm{D} \gg |\Delta_{12}^-|$ for all $n\geq 5$. As $\Delta_{12}^-$ couples $\ket{10}^\mathrm{L}$ and $\ket{11}^\mathrm{L}$ also to states outside the DFS, the fidelity shows a similar behaviour as the Pauli-X gate (see figure \ref{fig3}(f) for example with $N=4$), i.e., there is an optimal  $\Delta_{12}^-$ that sets the maximum fidelity.

The infidelity can be approximated similarly to equation \ref{eq:FidAppPh}, i.e., after a controlled Pauli-$Z$ gate ($|\Delta_{12}^-| T/\sqrt{2} = \pi$) acting on the state $\left( \ket{10+11}^\mathrm{L} \right)/\sqrt{2}$ can be approximated by
\begin{equation}
	1-F \approx \frac{3 \pi}{2 \sqrt{2} |\Delta_{12}^-|} 
		\left(\Gamma^* + \frac{3}{4} \frac{| \Delta_{12}^- |}{\Gamma_{1\mathrm{D}}} \right),
\end{equation}
which attains its minimal value, $3 \pi/\sqrt{2 P_\oned}$, for $|\Delta_{12}^-| = \sqrt{4 \Gamma^* \Gamma_\oned/3}$. As for the single qubit gates, the infidelity scales with $P_\oned^{-1/2}$, shown in blue circles figure \ref{fig4}.

\subsection{Summary of analysis}

\begin{figure}[tb]
	\centering
	\includegraphics[width=0.49\textwidth]{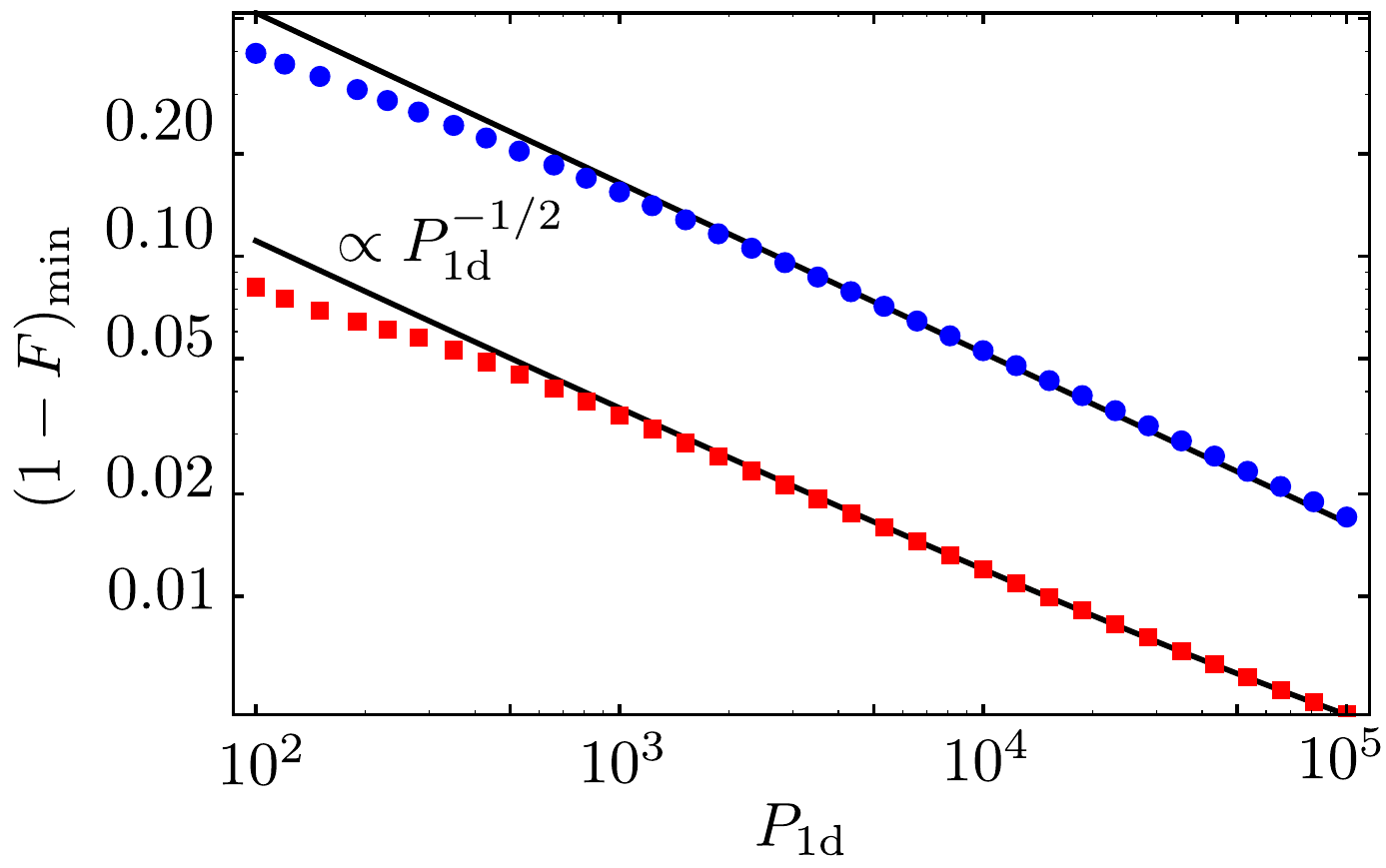}
	\caption{Scaling behaviour of the minimal infidelity for the Pauli-X (red squares) and controlled-$(-Z)$ gates (blue circles) with the same parameters as in figure \ref{fig3}e-f. The scaling fits well with the scaling $P_\oned^{-1/2}$ (black line) for large enough values of the Purcell Factor $P_\oned$. For the phase shift gate the infidelity is arbitrarily close to $0$ in the ideal case.} \label{fig4}
\end{figure}

Summing up, from the explicit analysis with two and four TLS, we have shown both numerically and analytically that both the single-qubit rotations and the control (-Z) gate show a scaling of the infidelity as $P_\oned^{-1/2}$ (see figure \ref{fig4}). Only for small values of the Purcell Factor $P_\oned$ does the minimal infidelity deviate slightly from the theoretical analysis because the hierarchy $\Gamma^* \ll \Omega_{12}^-, \Delta_{12}^- \ll \Gamma_\oned$ is no longer well satisfied.

Moreover, in the $N=4$ case, we also showed how to deal with the errors that come from the larger size of the DFS with respect to the computational one. For single qubit rotations in a system of four emitters, the choice $ |\Omega_{12}^-|, |\Delta_{12}^+| \ll |\Delta_{34}^+| \ll \Gamma_\oned$ ensures that the dynamics can be restricted to two atoms. In the extreme case where $|\Delta_{34}^+| \gg \Gamma_\oned$ the perturbation analysis is no longer valid. However, in this case the levels are so strongly shifted, that they are decoupled from the collective dissipation, so that the system can be described as a system of only two emitters. The same is true, if the emitters can be completely decoupled from the waveguide by other means available in a particular implementation.
For more atoms the same arguments hold as the second order correction introduced from deviations from Zeno dynamics satisfies
\begin{equation}
	\|\PP \mathcal{L}_{\mathrm{pert}} \QQ \frac{1}{\mathcal{L}_\DD}\QQ \mathcal{L}_{\mathrm{pert}}  \PP \rho \| 
	\ll \frac{\| \mathcal{L}_{\mathrm{pert}}  \|^2}{\Gamma_\oned},
\end{equation}
where $\mathcal{L}_{\mathrm{pert}}$ is the perturbation to the purely collective decay and $\| \cdot \|$ denotes the maximum norm. This is independent of the atom number $N$, because $\| \mathcal{L}_{\mathrm{pert}}  \|$ does not increase with the number of atoms for one and two-qubit gates. So there is an upper limit on the second order correction, which leads to the $P_\oned^{-1/2}$-scaling.
Finally, the error rate stemming from spontaneous emission of the logical states $\ket{1}_L=\ket{A}$ is proportional to the number of excited states in the system. Therefore, the gate fidelity does depend on the full state, and can be upper bounded by considering the worst-case state, that is the state with $\ket{1}_L$ in all other computational qubits, which indeed will depend on the atom number.

\section{Further error analysis: finite propagation length of 1d modes.}\label{sec5}

For completeness, it is interesting to consider another source of error that may be very relevant for some implementations with short propagation lengths, e.g., plasmonic waveguides \cite{chang07a,dzsotjan10a,gonzaleztudela11a,bermudez15a}. The finite propagation length enters into the decay matrix \cite{gonzaleztudela11a} as
\begin{equation}
	\Gamma_{n,m} = \frac{\Gamma_\oned}{2} \ee^{\ii q(\omega_\mathrm{a}) |z_n - z_m|}  
		\ee^{- |z_n - z_m|/L_\mathrm{prop}}
	= \frac{\Gamma_\oned}{2} \ee^{- x |n-m|},
\end{equation}
if the atoms are equidistantly placed a multiple of a wavelength apart, $d$ and where we introduced $x = d/L_\mathrm{prop}$ as the perturbation parameter. For simplicity, we restrict our discussion to the case with $N=2$, where analytical expressions can be obtained. In that situation, the finite propagation length only leads to the replacements 
\begin{eqnarray}
	\Gamma^* \rightarrow& \Gamma^* + \Gamma_\oned (1-\ee^{-x}) \approx \Gamma^* + \Gamma_\oned x,  \\
	2 \Gamma_\oned \rightarrow& \Gamma_\oned \left(2-(1-\ee^{-x})\right) \approx \Gamma_\oned \left(2-x \right)\,,
\end{eqnarray}
when $x\ll 1$. Therefore, the scaling of the infidelity is then given by
\begin{equation}
	1-F 
	\propto \sqrt{\frac{\Gamma^* + \Gamma_\oned x}{\Gamma_\oned \left(2-x \right)}} 
	\approx P_\oned^{-1/2} + \frac{1}{2} P_\oned^{1/2} x,
\end{equation}
which scales as $1/\sqrt{P_{\oned}}$ as long as $x P_\oned \ll 1$,that is, that the distance between neighbouring emitters satisfies $d=|z_n-z_{n+1}|\ll L_\mathrm{prop} P_\oned^{-1}$. For more atoms, it is difficult to obtain the analytical scaling as the superradiant state is not an eigenstate of the modified decay matrix and thus the DFS states change as well. However, because the finite propagation length enters as $\rme^{-|z_m -z_n|/L_\mathrm{prop}} \approx 1 - |z_m -z_n|/L_\mathrm{prop}$ it can be treated as a perturbation to the Liouvillian of equation \ref{Dicke} that will be kept small as long as $Nd\ll L_{\mathrm{prop}}$ \cite{gonzaleztudela15a}.  

Depending on the particular implementation other errors have to be considered, e.g., for atoms trapped close to a dielectric waveguide the separation condition $|z_n-z_m| = n 2\pi /q(\omega_\aa)$ might not be satisfied exactly or because its position is changing over time due to atomic motion. However, its main effect can be approximated as an effective increase of $\Gamma^*$ that is small with current state of the art parameters for photonic crystal waveguides, as discussed in reference \cite{gonzaleztudela15a}.

\section{Comparison to Three-Level Atoms} \label{sec6}

The use of the DFS of atomic $\Lambda$-systems in cavity QED setups has already been considered in detail in the literature \cite{beige00a,beige00b,tregenna02a}. In that case, a three-level system with a $\Lambda$-type level structure is used to define a logical qubit in the two metastable states $\ket{0}$ and $\ket{1}$. The excited state $\ket{e}$ decays to one of the metastable states, say $\ket{1}$. When two atoms are inside the cavity an additional decoherence-free state emerges, i.e.,  $\left(\ket{1e}-\ket{e1}\right)/\sqrt{2}$, that can be used to define a CNOT gate in the so-called bad-cavity limit, where  the atom-cavity coupling ($g$) is smaller than the cavity losses $\kappa$, but the decay into the cavity ($g^2/\kappa$) is still bigger than into the rest of the decay channels ($\Gamma^*$). The ratio between the good/bad processes is the so-called cooperativity $C=\frac{g^2}{\kappa \Gamma^*}$, which therefore plays a similar role as $P_{\mathrm{1d}}$ in our proposal. The errors in the CNOT gate come both from $\Gamma^*$, and from deviations from the Zeno Hamiltonian, giving rise to an optimal infidelity proportional to $1/\sqrt{C}$, which is similar to the one that we found using only TLSs.

We note that using TLS the computational qubits have a finite lifetime compared to the implementations using atomic metastable states. However, i) there are situations in which one would like to use gates to build a given atomic state within the DFS in order to map it immediately into a photonic state in the waveguide \cite{gonzaleztudela15a} such that long lifetimes are not required; ii) some of the implementations have extremely long-lived qubits, e.g.~superconducting systems \cite{mlynek14a}. iii) Moreover, if $\Lambda$ schemes are available, as in atoms, we can also implement our single and two-qubit gates with metastable states with the equivalence shown in figure \ref{fig1}c. In that case, our proposal just constitutes a complementary way of doing universal quantum computation within DFS.

\section{Conclusion \& Outlook.}

Summing up, we have shown how to implement a universal set of quantum gates using the decoherence-free subspaces appearing within TLS interacting with one-dimensional photon-like reservoirs. We have given an explicit construction of single and two-qubit gates for logical qubits defined in the DFS and analyzed possible sources of errors such as spontaneous emission to other modes, coupling to states outside of the DFS, imperfect addressing and finite propagation lengths. Through both analytical and numerical analysis, we have shown the fidelities of the gates scale generally with $(1-F)_\mathrm{min}\propto P_\oned^{-1/2}$, analogous to the one using $\Lambda$ schemes  \cite{beige00a,beige00b,tregenna02a}. Thus, this work widens up the zoology of quantum emitters that can be used to implement quantum gates within waveguide QED setups. An interesting outlook for the application of these gates is to use them for generating entangled states of many emitters within the DFS, which afterwards can be mapped into waveguide multiphoton states in a very efficient way \cite{gonzaleztudela15a}.

\ack
We gratefully acknowledge discussions with I. Cirac. The work of AGT and VP was funded by the European Union integrated project \emph{Simulators and Interfaces with Quantum Systems} (SIQS). AGT also acknowledges support from Alexander Von Humboldt Foundation and Intra-European Marie-Curie Fellowship NanoQuIS (625955). HJK acknowledges funding by the Institute of Quantum Information and Matter, a National Science Fundation (NSF) Physics Frontier Center with support of the Moore Foundation, by the Air Force Office of Scientific Research, Quantum Memories in Photon-Atomic-Solid State Systems (QuMPASS) Multidisciplinary University Research Initiative (MURI), by the  Department of Defense National Security Science and Engineering Faculty Fellows (DoD NSSEFF) program, by NSF PHY1205729 and support as a Max Planck Institute for Quantum Optics Distinguished Scholar.

\appendix
\setcounter{section}{1}
\setcounter{equation}{0}

\section*{Appendix A: Second order Liouvillian perturbation theory.\label{sec:apA}}

When the system evolves under a very strong collective decay, the driving term $H_\mathrm{las}$ and the decay into other bath modes may be treated as a perturbation to the collective dissipation given by $\mathcal{L}_\mathrm{D}$ \cite{reiter12a,kessler12b}. In order to describe these perturbations as generally as possible, we denote them by a Liouvillian $\mathcal{L}_\mathrm{pert}$, and assume that it has a relevant timescale $\tau$. If the timescale satisfies, $\tau\gg 1/\Gamma_{\oned}$, the dynamics of the atomic system can be formally projected into the DFS of the Liovillian ${\cal L}_\DD$, by using a projector operator $\mathbb{P}$ satisfying: $\mathbb{P}{\cal L}_\DD={\cal L}_\DD \mathbb{P}=0$. This projector can be found via the right (left) eigenoperators $\rho_{ij}$ ($\chi_{ij}$) corresponding to the eigenvalue 0 of $\mathcal{L}_\mathrm{D}$, which are combined to 
\begin{equation}
	\mathbb{P}\rho = \sum_{i,j} \rho_{ij} \langle \chi_{ij}, \rho \rangle,
\end{equation}
where $\langle A,B \rangle = \Tr \left(A^\dagger B\right)$ is the inner product on the space of density matrices.
The orthogonal eigenoperators are indexed such that $ \langle \chi_{ij}, \rho_{kl} \rangle = \delta_{i,k} \delta_{j,l} = \langle \rho_{ij}, \rho_{kl} \rangle$.
We also define the orthogonal part of $\mathbb{P}$, $\mathbb{Q}=1-\mathbb{P}$. The left eigenoperators can be derived from the right ones by 
\begin{equation}
	\chi_{ij} = \rho_{ij} + \alpha_{ij}^{(1)} S_{eg} \rho_{ij} S_{ge} + \alpha_{ij}^{(2)} S_{eg} S_{eg} \rho_{ij} S_{ge} S_{ge} + \dots,
\end{equation}
where the coefficients $\alpha_{ij}^{(n)}$ are determined by $\langle \chi_{ij}, \mathcal{L}_\DD \rho \rangle = 0$. With this choice, the projector is independent of the choice of $\rho_{ij}$ and hence, one can choose $\rho_{ij} = \ket{\mathrm{d}_i} \bra{\mathrm{d}_j}$, where $\ket{\mathrm{d}_i}$ are orthonormal states from the DFS. Using these projectors, one can formally integrate out the fast dynamics outside the DFS, described by $\QQ \rho$:
\numparts
\begin{eqnarray}
	\frac{\dd}{\dd t} \QQ \rho 
		&= \mathbb{Q}\ \left( \mathcal{L}_\DD + \mathcal{L}_\mathrm{pert} \right) \mathbb{Q} \rho + \mathbb{Q} \mathcal{L}_\mathrm{pert} \mathbb{P} \rho \, , \\
	\mathbb{Q} \rho(t) 
		&= \int_0^t \rmd \tau\ \mathrm{exp}\left[ \mathbb{Q} (\mathcal{L}_\DD+ \mathcal{L}_\mathrm{pert}) \mathbb{Q} (t-\tau) \right]\mathbb{Q}  \mathcal{L}_\mathrm{pert} \mathbb{P} \rho(\tau) \nonumber \\
		& \approx \mathbb{Q} (-\mathcal{L}_\DD^{-1} ) \mathbb{Q} \mathcal{L}_\mathrm{pert} \mathbb{P} \rho + \mathcal{O}(\tau^{-2} / \Gamma_\oned^2),
	\label{eq:Qrho}
\end{eqnarray}
\endnumparts
where the last approximation is obtained by i) applying a Markov approximation $\rho(\tau) \approx \rho(t)$ in the integral; ii) neglecting terms of higher order in $\tau^{-1}/\Gamma_\oned$; and iii) extending the integral to infinity.

Plugging this into the equation for the DFS-part of the state, that is
\begin{equation}
	\label{eqA:meq}
	\frac{\dd}{\dd t} \PP \rho 
	= \PP \mathcal{L}_\mathrm{pert} \PP \rho + \PP \mathcal{L}_\mathrm{pert} \QQ \rho
	= \mathcal{L}_\mathrm{eff} \PP \rho ,
\end{equation}
yields an effective Liouvillian of the atomic system within the DFS given (up to second order in $\tau^{-1}/\Gamma_\oned$) by

\begin{equation}
	\mathcal{L}_\mathrm{eff} 
	= \PP  \mathcal{L}_\mathrm{pert} \PP 
	+\PP  \mathcal{L}_\mathrm{pert} \QQ \frac{1}{{-\cal L}_\DD}\QQ  \mathcal{L}_\mathrm{pert} \PP
	+\mathrm{O}\left(\tau^{-3}/\Gamma_\oned^2 \right)\,.
\end{equation}

The first order of this Liouvillian, i.e., $\PP  \mathcal{L}_\mathrm{pert} \PP$, is the effective evolution induced within the DFS induced by the strong collective dissipation. This is commonly referred to as the ideal Quantum Zeno dynamics \cite{zanardi97a,lidar98a,facchi02a} as it can be understood as the effective dynamics enforced by the continuous monitoring of the atomic system due to the interaction of the waveguide modes. The second order term stems from slight population of (super)radiant modes that generates some corrections on the ideal Quantum Zeno dynamics.

It is instructive to write the effective master equation derived in Eq.~\ref{eqA:meq} in a form that separates the non-hermitian evolution dynamics and the contribution coming from quantum jump processes. For our particular situation, considering the perturbation of $\mathcal{L}_*$ and $H_\mathrm{las}$ as defined in the main text, it can be shown after some algebra that
\begin{equation}
	\label{eqA:effmas}
	\dot{\rho} 
	= - \ii \left[H_\mathrm{eff},\rho \right] 
	+ \mathbb{P} \mathcal{L}_* \rho 
	+ \mathbb{P} \left( o_1 \rho o_2^\dagger + o_2 \rho o_1^\dagger \right) 
	- o_2^\dagger o_1 \rho - \rho o_2^\dagger o_1,
\end{equation}
where $H_\mathrm{eff}=\mathcal{P} H_\mathrm{las} \mathcal{P}$, where we used the projection $\mathcal{P}$ [$\mathcal{Q}$] for pure states inside [outside] of the DFS as defined in the main text.
Furthermore, $o_1 = \mathcal{Q} \left(\frac{\Gamma_\oned}{2} S^+  S^-\right)^{-1} \mathcal{Q} H_\mathrm{las} \mathcal{P}$, $o_2 =\mathcal{Q} H_\mathrm{las} \mathcal{P}$ can be obtained by noting that the second order term reduces to simple matrix multiplication in the corresponding subspace because $\mathbb{P} \mathcal{P} A \mathcal{Q} = 0 = \mathbb{P} \mathcal{Q} A \mathcal{P}$ for all operators $A$.\footnote{We want to note that if it was possible to detect the photons emitted to the waveguide and post-select, the term $\mathbb{P} \left( o_1 \rho o_2^\dagger + o_2 \rho o_1^\dagger \right)$ would vanish and the system would be described by pure states except for the spontaneous emission into non-guided modes.}
Although this does not look like a Liouvillian in Lindblad form, it is trace-preserving, as
\begin{equation}
	\Tr \mathbb{P} A 
	= \sum_{i,j} \Tr \rho_{ij} \langle \chi_{ij}, A \rangle 
	= \sum_i \langle \chi_{ii}, A \rangle 
	= \Tr A,
\end{equation}
because $\sum_{i} \chi_{ii} = \mathbf{1}$. From Eq.~\ref{eqA:effmas}, it is straightforward to define a non-Hermitian Hamiltonian from the above master equation, that is
\begin{equation}
	H_\mathrm{nh} 
	= \mathcal{P} \left( H_\mathrm{eff} 
	- \ii \frac{\Gamma^*}{2} \sum_n \sigma^n_{eg} \sigma_{ge}^n  - \ii o_2^\dagger o_1 \right)\mathcal{P},
\end{equation}
which describes the no-jump evolution, and that we use to get the analytical estimations of the infidelity.

\section*{References}
\bibliographystyle{unsrt}
\bibliography{Sci,books}

\begin{thebibliography}{10}

\bibitem{joannopoulos_book95a}
J.~D. Joannopoulos, R.~D. Meade, and J.~N. Winn.
\newblock {\em {Photonic Crystals: Molding the Flow of Light}}.
\newblock Princeton University Press, 1995.

\bibitem{laucht12a}
A.~Laucht, S.~P{\"u}tz, T.~G{\"u}nthner, N.~Hauke, R.~Saive,
  S.~Fr{\'e}d{\'e}rick, M.~Bichler, M.-C. Amann, A.~W. Holleitner, M.~Kaniber,
  and J.~J. Finley.
\newblock {A Waveguide-Coupled On-Chip Single-Photon Source}.
\newblock {\em Phys. Rev. X}, 2:011014, Mar 2012.

\bibitem{goban13a}
A.~Goban, C.-L. Hung, S.-P Yu, J.D. Hood, J.A. Muniz, J.H. Lee, M.J. Martin,
  A.C. McClung, K.S. Choi, D.E. Chang, O.~Painter, and H.J. Kimble.
\newblock Atom-light interactions in photonic crystals.
\newblock {\em Nat. Commun.}, 5:3808, 2014.

\bibitem{yu14a}
S-P Yu, JD~Hood, JA~Muniz, MJ~Martin, Richard Norte, C-L Hung, Se{\'a}n~M
  Meenehan, Justin~D Cohen, Oskar Painter, and HJ~Kimble.
\newblock Nanowire photonic crystal waveguides for single-atom trapping and
  strong light-matter interactions.
\newblock {\em Appl. Phys. Lett.}, 104(11):111103, 2014.

\bibitem{tiecke14a}
TG~Tiecke, JD~Thompson, NP~de~Leon, LR~Liu, V~Vuleti{\'c}, and MD~Lukin.
\newblock Nanophotonic quantum phase switch with a single atom.
\newblock {\em Nature}, 508(7495):241--244, 2014.

\bibitem{sollner14a}
Immo S{\"o}llner, Sahand Mahmoodian, Alisa Javadi, and Peter Lodahl.
\newblock A chiral spin-photon interface for scalable on-chip
  quantum-information processing.
\newblock {\em arXiv:1406.4295}.

\bibitem{arcari14a}
M.~Arcari, I.~S\"ollner, A.~Javadi, S.~Lindskov~Hansen, S.~Mahmoodian, J.~Liu,
  H.~Thyrrestrup, E.~H. Lee, J.~D. Song, S.~Stobbe, and P.~Lodahl.
\newblock Near-unity coupling efficiency of a quantum emitter to a photonic
  crystal waveguide.
\newblock {\em Phys. Rev. Lett.}, 113:093603, Aug 2014.

\bibitem{goban15a}
A.~Goban, C.-L. Hung, J.~D. Hood, S.-P. Yu, J.~A. Muniz, O.~Painter, and H.~J.
  Kimble.
\newblock Superradiance for atoms trapped along a photonic crystal waveguide.
\newblock {\em Phys. Rev. Lett.}, 115:063601, Aug 2015.

\bibitem{young15a}
A.~B. Young, A.~C.~T. Thijssen, D.~M. Beggs, P.~Androvitsaneas, L.~Kuipers,
  J.~G. Rarity, S.~Hughes, and R.~Oulton.
\newblock Polarization engineering in photonic crystal waveguides for
  spin-photon entanglers.
\newblock {\em Phys. Rev. Lett.}, 115:153901, Oct 2015.

\bibitem{vetsch10a}
E~Vetsch, D~Reitz, G~Sagu{\'e}, R~Schmidt, ST~Dawkins, and A~Rauschenbeutel.
\newblock Optical interface created by laser-cooled atoms trapped in the
  evanescent field surrounding an optical nanofiber.
\newblock {\em Phys. Rev. Lett.}, 104(20):203603, 2010.

\bibitem{goban12a}
A.~Goban, K.~S. Choi, D.~J. Alton, D.~Ding, C.~Lacro{\^u}te, M.~Pototschnig,
  T.~Thiele, N.~P. Stern, and H.~J. Kimble.
\newblock {Demonstration of a State-Insensitive, Compensated Nanofiber Trap}.
\newblock {\em Phys. Rev. Lett.}, 109:033603, 2012.

\bibitem{mitsch14a}
R~Mitsch, C~Sayrin, B~Albrecht, P~Schneeweiss, and A~Rauschenbeutel.
\newblock Quantum state-controlled directional spontaneous emission of photons
  into a nanophotonic waveguide.
\newblock {\em Nature Commun.}, 5:5713, 2014.

\bibitem{petersen14a}
Jan Petersen, J{\"u}rgen Volz, and Arno Rauschenbeutel.
\newblock Chiral nanophotonic waveguide interface based on spin-orbit
  interaction of light.
\newblock {\em Science}, 346(6205):67--71, 2014.

\bibitem{beguin14a}
J.-B. B\'eguin, E.~M. Bookjans, S.~L. Christensen, H.~L. S\o{}rensen, J.~H.
  M\"uller, E.~S. Polzik, and J.~Appel.
\newblock Generation and detection of a sub-poissonian atom number distribution
  in a one-dimensional optical lattice.
\newblock {\em Phys. Rev. Lett.}, 113:263603, Dec 2014.

\bibitem{chang07a}
D.~E. Chang, A.~S. S{\o}rensen, P.~R. Hemmer, and M.~D. Lukin.
\newblock {Strong coupling of single emitters to surface plasmons}.
\newblock {\em Phys. Rev. B}, 76(3):035420, Jul 2007.

\bibitem{dzsotjan10a}
David Dzsotjan, Anders~S. S{\o}rensen, and Michael Fleischhauer.
\newblock {Quantum emitters coupled to surface plasmons of a nanowire: A
  Green's function approach}.
\newblock {\em Phys. Rev. B}, 82:075427, Aug 2010.

\bibitem{gonzaleztudela11a}
A.~Gonzalez-Tudela, D.~Martin-Cano, E.~Moreno, L.~Martin-Moreno, C.~Tejedor,
  and F.~J. Garcia-Vidal.
\newblock {Entanglement of Two Qubits Mediated by One-Dimensional Plasmonic
  Waveguides}.
\newblock {\em Phys. Rev. Lett.}, 106(2):020501, Jan 2011.

\bibitem{bermudez15a}
Esteban Berm{\'u}dez-Ure{\~n}a, Carlos Gonzalez-Ballestero, Michael Geiselmann,
  Renaud Marty, Ilya~P Radko, Tobias Holmgaard, Yury Alaverdyan, Esteban
  Moreno, Francisco~J Garc{\'\i}a-Vidal, Sergey~I Bozhevolnyi, et~al.
\newblock Coupling of individual quantum emitters to channel plasmons.
\newblock {\em Nature Communications}, 6, 2015.

\bibitem{koppens11a}
Frank~HL Koppens, Darrick~E Chang, and F~Javier Garcia~de Abajo.
\newblock Graphene plasmonics: a platform for strong light--matter
  interactions.
\newblock {\em Nano Letters}, 11(8):3370--3377, 2011.

\bibitem{huidobro12a}
P.~A. Huidobro, A.~Y. Nikitin, C.~Gonz{\'a}lez-Ballestero,
  L.~Mart{\'\i}n-Moreno, and F.~J. Garc{\'\i}a-Vidal.
\newblock {Superradiance mediated by graphene surface plasmons}.
\newblock {\em Phys. Rev. B}, 85:155438, Apr 2012.

\bibitem{christensen11a}
Johan Christensen, Alejandro Manjavacas, Sukosin Thongrattanasiri, Frank~HL
  Koppens, and F~Javier García~de Abajo.
\newblock Graphene plasmon waveguiding and hybridization in individual and
  paired nanoribbons.
\newblock {\em ACS nano}, 6(1):431--440, 2011.

\bibitem{martinmoreno15a}
Luis Martin-Moreno, F~de~Abajo, and Francisco~J Garcia-Vidal.
\newblock Ultra-efficient coupling of a quantum emitter to the tunable guided
  plasmons of a carbon nanotube.
\newblock {\em arXiv:1502.02488}.

\bibitem{mlynek14a}
JA~Mlynek, AA~Abdumalikov, C~Eichler, and A~Wallraff.
\newblock Observation of dicke superradiance for two artificial atoms in a
  cavity with high decay rate.
\newblock {\em Nat. Communications}, 5:5186, 2014.

\bibitem{lehmberg70a}
R.~H. Lehmberg.
\newblock {Radiation from an $N$-Atom System. I. General Formalism}.
\newblock {\em Phys. Rev. A}, 2(3):883--888, Sep 1970.

\bibitem{lehmberg70b}
R.~H. Lehmberg.
\newblock {Radiation from an $N$-Atom System. II. Spontaneous Emission from a
  Pair of Atoms}.
\newblock {\em Phys. Rev. A}, 2:889--896, Sep 1970.

\bibitem{zanardi97a}
Paolo Zanardi and Mario Rasetti.
\newblock Noiseless quantum codes.
\newblock {\em Physical Review Letters}, 79(17):3306, 1997.

\bibitem{lidar98a}
D.~A. Lidar, I.~L. Chuang, and K.~B. Whaley.
\newblock {Decoherence-Free Subspaces for Quantum Computation}.
\newblock {\em {Phys. Rev. Lett.}}, 81:2594, 1998.

\bibitem{beige00a}
Almut Beige, Daniel Braun, Ben Tregenna, and Peter~L Knight.
\newblock Quantum computing using dissipation to remain in a decoherence-free
  subspace.
\newblock {\em Phys. Rev. Lett.}, 85(8):1762, 2000.

\bibitem{beige00b}
Almut Beige, Daniel Braun, and Peter~L Knight.
\newblock Driving atoms into decoherence-free states.
\newblock {\em New Journal of Physics}, 2(1):22, 2000.

\bibitem{tregenna02a}
Ben Tregenna, Almut Beige, and Peter~L Knight.
\newblock Quantum computing in a macroscopic dark period.
\newblock {\em Physical Review A}, 65(3):032305, 2002.

\bibitem{facchi02a}
P.~Facchi and S.~Pascazio.
\newblock Quantum zeno subspaces.
\newblock {\em Phys. Rev. Lett.}, 89:080401, Aug 2002.

\bibitem{chang12a}
DE~Chang, L~Jiang, AV~Gorshkov, and HJ~Kimble.
\newblock Cavity qed with atomic mirrors.
\newblock {\em New Journal of Physics}, 14(6):063003, 2012.

\bibitem{gardiner_book00a}
G.~W. Gardiner and P.~Zoller.
\newblock {\em {Quantum Noise}}.
\newblock Springer-Verlag, Berlin, 2nd edition, 2000.

\bibitem{dicke54a}
R.~H. Dicke.
\newblock {Coherence in Spontaneous Radiation Processes}.
\newblock {\em {Phys. Rev.}}, 93:99, 1954.

\bibitem{nielsen_book00a}
M.~A. Nielsen and I.~L. Chuang.
\newblock {\em {Quantum computation and quantum information}}.
\newblock {Cambridge University Press}, 2000.

\bibitem{gonzaleztudela15a}
A.~Gonz\'{a}lez-Tudela, V.~Paulisch, D.~E. Chang, H.~J. Kimble, and J.~I.
  Cirac.
\newblock {Deterministic generation of arbitrary photonic states assisted by
  dissipation}.
\newblock {\em {Phys. Rev. Lett.}}, 115:163603, Oct 2015.

\bibitem{reiter12a}
Florentin Reiter and Anders~S S{\o}rensen.
\newblock Effective operator formalism for open quantum systems.
\newblock {\em Physical Review A}, 85(3):032111, 2012.

\bibitem{kessler12b}
E.~M. Kessler.
\newblock {Generalized Schrieffer-Wolff formalism for dissipative systems}.
\newblock {\em Phys. Rev. A}, 86:012126, Jul 2012.

\end{thebibliography}

\end{document}